\documentclass{iaus}
\usepackage{graphicx}

\title[Kepler Circumbinary Planets] 
{Recent Kepler Results On Circumbinary Planets}

\author[Welsh, et al.]   
{William F. Welsh$^1$,
Jerome A. Orosz$^1$,
Joshua A. Carter$^2$,
Daniel C. Fabrycky$^3$
\and the {\it{Kepler}} Team}

\affiliation{
$^1$Astronomy Department, San Diego State University,
San Diego, CA 92182-1221 USA
\\ email: {\tt wwelsh@mail.sdsu.edu}
\\[\affilskip]
$^2$ Harvard-Smithsonian Center for Astrophysics, Cambridge, MA 02138 
USA
\\[\affilskip]
$^3$ Dept. of Astronomy and Astrophysics, University of 
California, Santa Cruz, CA 95064 USA; and
Department of Astronomy and Astrophysics, University of Chicago, 
5640 South Ellis Avenue, Chicago, IL 60637 USA
}

\pubyear{2013}
\volume{293}  
\pagerange{xxx--xxx}
\jname{Formation, detection, and characterization of extrasolar 
habitable planets}
\editors{Nader Haghighipour}
\begin{document}

\maketitle

\begin{abstract}
Ranked near the top of the long list of exciting discoveries 
made with NASA's {\it Kepler} photometer is the detection
of transiting circumbinary planets. In just over a year the
number of such planets went from zero to seven, including a 
multi-planet system with one of the planets in the habitable
zone (Kepler-47).
We are quickly learning to better detect and characterize these 
planets, including the recognition of their transit timing and 
duration variation ``smoking gun'' signature. Even with only 
a handful of such planets, 
some exciting trends are emerging.

\end{abstract}

\firstsection 
\section{Introduction}
The first transiting circumbinary planet, Kepler-16b, was 
announced in September of 2011
(\cite[Doyle et al.\ 2011]{Doyle2011}).
The fact that {\it transits} are present is crucial, in that they
unambiguously establish the presence of the circumbinary body.
In addition to being a key to circumbinary planet detection,
the transits further provide the information necessary to obtain
very precise stellar and planetary masses and radii, via the 
exact geometrical configuration necessary to produce the detailed
shape, duration, and timing of the transits.
The primary star's mass in Kepler-16 is known to better than 
0.5\%, and its radius to better than 0.2\%. Similarly, the 
secondary star (an M star with mass of only 0.20 M$_{\odot}$) has 
uncertainties in its mass and radius of less than 0.33\% and 
0.26\%. The planet's mass and radius are
likewise known to remarkable precision, 4.8\% and 0.34\%, 
respectively. Also, the primary star's rotation axis has been measured
to be aligned with the binary's orbital axis to within 2.4 degrees
(\cite[Winn et al.\ 2011]{Winn2011}).
The circumbinary configuration provides a boon to stellar 
and exoplanet astrophysics.

But despite these impressive results, 
many questions remained unanswered: e.g.\
What kinds of planetary orbits are possible (what periods and 
eccentricities)?
What kinds of stellar orbits are possible? 
Is there a preferred stellar mass ratio?
What planetary radii, masses, and temperatures are possible?
And perhaps most importantly, was Kepler-16 just a lucky accidental 
quirk, or are circumbinary planets common?

In the short time since Kepler-16 was discovered,
five\footnote{
The sixth system (seventh planet) was discovered and announced 
by the Planet Hunters after the IAU Symposium by
\cite[Schwamb, et al.\ (2012)]{Schwamb2013},
and was also independently discovered by
\cite[Kostov, et al.\ (2012)]{Kostov2013}.
}
additional transiting circumbinary planet systems have been 
confirmed, including the Kepler-47 system 
(\cite[Orosz, et al.\ 2012b]{Orosz2012b})
discussed below.
Moreover, several candidate systems are under investigation.
These discoveries establish that Kepler-16 was not a rare
oddity and that the circumbinary planets are indeed a new 
class of planet
(\cite[Welsh, et al.\ 2012]{Welsh2012}).
These discoveries are spurring on investigations of
circumbinary planet formation and stability (e.g.
\cite[Meschiari (2012)]{Meschiari2012},
\cite[Paardekooper, et al.\ (2012)]{Paardekooper2012},
\cite[Youdin, et al.\ (2012)]{2012},
\cite[Rafikov (2013)]{Rafikov2013}).
In the following section we discuss the multi-planet Kepler-47
system and the rapidly developing picture of the nature of 
circumbinary planets.

\section{The Circumbinary Planet ``Smoking Gun''}
Transits\footnote{For clarity, we use the terminology ``primary''
and ``secondary'' eclipse for the stars, ``primary transit'' when 
the planet transits the primary star, ``secondary transit'' when 
the planet transits the secondary star, ``primary occultation'' 
when the planet is occulted by the primary star, etc.}
in a circumbinary system are harder to detect than transits in a 
single-star system for a variety of reasons.
The eclipses are usually very much stronger than the transits, and 
their presence tends to swamp out or mask the transit signal.
But the more challenging issue is that the transits are neither
periodic nor of equal duration. A single planet orbiting a single
star will exhibit strictly periodic transits. Searches based on
a periodic signal can be highly efficient at detecting the planet.
If there are other planets in the system that strongly 
gravitationally interact with the transiting planet, the times of 
transits are no longer periodic, but can vary by a few tens of 
minutes to a few hours, typically,
e.g. Kepler-9 b and c 
(\cite[Holman et al.\ 2010]{Holman2010}).
These are the well-known transit timing variations (TTVs) that can
be used for planet confirmation and mass determination.
In contrast, for a circumbinary planet, the transit period is
very far from being strictly periodic. For example, in Kepler-16,
the third transit across the primary star occurs 8.8 {\it{days}}
(days, not hours) early compared to the expected time based on the 
first two transits 
(\cite[Doyle et al.\ 2011]{Doyle2011}).
The reason for the changing period is that the primary star is not 
stationary with respect to the planet: it orbits the barycenter of 
the binary star system. Sometimes the transits occur when the star
is before stellar conjunction, and sometime transits occur after 
conjunction. 
If ${\rm{ M_{planet} \ll M_{2} \ll M_{1} }}$, the geometrical
configuration-induced TTVs will be $\sim$ zero since 
the primary is always near the system barycenter. Equal mass stars 
will result in the largest TTVs, and in a configuration with the 
planet and stars coplanar and on circular orbits, the deviation of 
the primary transit times from a linear ephemeris is a 
minimum of $\pm$ $\sim 0.11$ times the orbital period of the binary 
stars. 
For more general orbital orientations and eccentricities the TTVs 
can be even larger; and transits of the secondary star will have 
yet larger TTVs, e.g.,
$\pm \sim 12$ days for Kepler-16b secondary transits.
Obviously the TTVs will limit the success of planet searches based 
on a periodic signal.

The other effect that is important is that the widths of the 
transits can vary by very substantial amounts, 
yielding transit duration variations (TDVs). 
The varying transit widths are often 
readily detectable by eye in the light curves. 
This is again due to the ``moving target'' effect.
Assuming the planet orbits in the same sense and in nearly the same 
plane as the binary, the transit duration is set by the size of the 
star (and planet) and the relative velocity of the star and planet 
(i.e. their transverse velocity). Near binary phase zero (defined by 
the primary eclipse), the star and planet are moving in the opposite 
direction, and the transits are short in duration.
Near secondary eclipse, the star and planet are moving in the same
direction and the transits can be very long. 
Near the quadrature points of the binary orbit, the 
transits are ``normal'' duration. The effect is strong:
{\it eclipse durations can vary by factors of several,} in some
cases varying from a few hours to a few days.
This TDV phenomenon for planets orbiting binary stars was noted 
as early as 1996 
(\cite[Jenkins et al.\ 1996]{Jenkins1996},
and in particular see their figures 5 and 6), 
several years before the first transiting planet was discovered.

Because the transit widths vary so much, search algorithms 
that simply fold transits on a linear ephemeris are not optimal 
for circumbinary planet detection. 
Fortunately, sophisticated methods to cope with the TTV and TDVs do 
exist, notably the ``TDA matched filter method''  
(\cite[Jenkins et al.\ 1996)]{Jenkins1996}, the 
``CB-BLS algorithm'' (\cite[Ofir 2008)]{Ofir2008},
and most recently the ``QATS'' algorithm in EB mode
(\cite[Carter \& Agol 2013)]{Carter2013}. 
As the search for transiting circumbinary 
planets pushes to super-Earth-size planets and smaller (so that 
individual transits can not be detected visually), such automated 
search methods will be of great value.

While these TTVs and TDVs make circumbinary planet hunting difficult, 
they do have a very important positive consequence. 
When a candidate transiting circumbinary planet is found, it is 
very easy to confirm or refute it is a physically bound body. 
The binary star phase-specific TTV and TDVs provide a 
`smoking gun'; no other known astrophysical phenomenon, such
as a background eclipsing binary, can mimic such variations.

If the planet is massive enough to measurably alter the stars' orbits, 
stellar eclipse timing variations (ETVs) will be apparent. 
Because eclipses are generally much deeper than transits, eclipse 
times can usually be measured with much higher precision than transit
times, allowing for detection of timing deviations as small as a 
few seconds. The most readily measurable effect of a planet on the 
binary stars is the precession of the stellar orbits that results in 
a difference between the orbital period as defined by the primary 
eclipses versus the orbital period as defined by the secondary 
eclipses. The observed-minus-computed times of the primary eclipses
and secondary eclipses then show a diverging trend with time,
e.g., see \cite[Welsh, et al.\ (2012)]{Welsh2012}.
In addition to this dynamical perturbation, a light-travel time 
effect (LITE) can also be present and help constrain the mass of the 
planet relative to the binary. Thus the moving target aspect of 
circumbinary planet
host stars allows for a rock-solid confirmation of the planets based 
on the presence of:
1)~transits; 2)~very large TTVs; 3)~very large TDVs.
Also very helpful for a full solution of the system parameters are
4)~ETVs and 5)~LITE.

\begin{figure}[h!]
\begin{center}
 \includegraphics[width=3.4in]{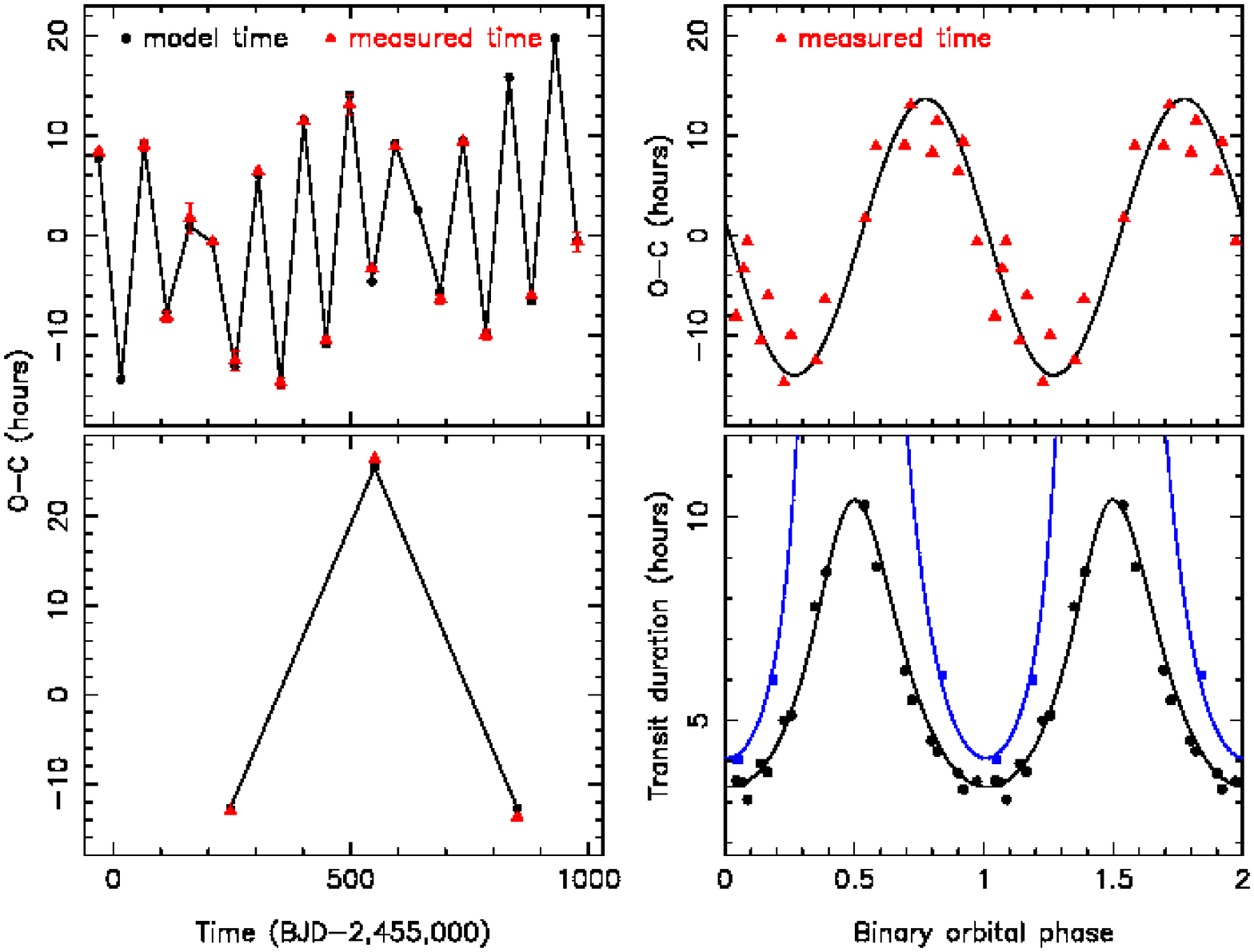} 
 \caption{
{\it left:} Transit timing variations and transit duration variations 
for Kepler-47 b (top) and c (bottom).
{\it right:} TTVs for planet b and TDVs for both planets phase-folded 
onto the binary star's 7.4 day orbital period.
}
\label{fig1}
\end{center}
\end{figure}

\section{Kepler-47}
Visual inspection of the light curve of the eclipsing binary
KIC 10020423 (KOI-3154) revealed transits of two candidate 
circumbinary planets. The outer candidate, ``c'', has a transit depth
of $\sim$0.2\% and was readily detectable, but the inner candidate 
``b'' was harder to find because of its shallower $\sim$0.1\% 
transit depth. The system is faint, with Kp=15.2 mag, and thus noisier 
than other {\it Kepler} circumbinary systems. In addition,
variations of amplitude 2-4\% in the light curve due to 
starspots made detection considerably more difficult. Three 
transits of candidate c were discovered, and using a preliminary
model of the system as a guide, 18 transits of candidate b were 
eventually found. 
A modified version of the photometric-dynamical code developed for 
the previous {\it Kepler} circumbinary planets was used to model 
the light curve and radial velocities. The code assumes spherical 
bodies moving under the influence of Newtonian gravity. 
Since the planets are not sufficiently massive to noticeably perturb 
the binary stars nor each other, they were treated as
massless bodies to greatly reduce the computation time.
Details of the photometric-dynamical code are provided in the 
Supplementary Material of 
\cite[Orosz, et al.\ (2012b)]{Orosz2012b}.
While the planets do not induce measurable dynamical effects, 
the ``moving target effect'' causes 
planet b to exhibit peak-to-peak TTVs of $\sim$28 hrs 
and TDVs of $\sim$7.2 hrs.
Planet c has observed peak-to-peak TTVs of 39.7 hr and TDVs of 
2.4 hrs.
The best-fit photometric-dynamical model gives an excellent match 
to the data, confirming these as circumbinary planets.

The primary eclipses are $\sim$15\% deep while the (total) secondary 
eclipses are only $\sim$0.6\% deep. Combining the {\it Kepler}
photometry with radial velocities obtained at the McDonald  
Observatory allowed a partial solution of the binary star parameters.
A full solution was not possible because only the primary star's
radial velocities were detected in the spectra. However, tight
constraints on the primary star's location and motion in its orbit 
are placed by the planetary transits. This provided the mass ratio 
of the system, and thus the full set of binary star system parameters.
The primary star is quite Sun-like, with a mass of $1.04 \pm 0.06$ 
M$_{\odot}$, radius $0.96 \pm 0.02$ R$_{\odot}$, and temperature of 
$5636 \pm 100$~K.
The secondary is a diminutive star of mass 0.36 M$_{\odot}$ and 
radius $0.351 \pm 0.006$ R$_{\odot}$, and emitting only 0.57\% the 
flux of the primary in the {\it Kepler} bandpass. With the stellar 
radii known, the planetary radii can then be determined. Planet b 
has a radius of 3.0 R$_{\rm{Earth}}$, making it the smallest 
transiting circumbinary planet yet discovered. Planet c has a 
radius of 4.6 R$_{\rm{Earth}}$, or 
$\sim$1.2 R$_{\rm{Neptune}}$.
While the masses are currently below our detection threshold,
3-$\sigma$ upper limits of $<2$ M$_{\rm{Jup}}$ and 
$<28$ M$_{\rm{Jup}}$ 
can be placed, based on the lack of measurable ETVs and the lack of 
LITE of the binary. Because of their small radii, the masses are 
likely to be very much smaller than these upper limits.

\begin{figure}[ht]
\begin{center}
\includegraphics[width=3.8in]{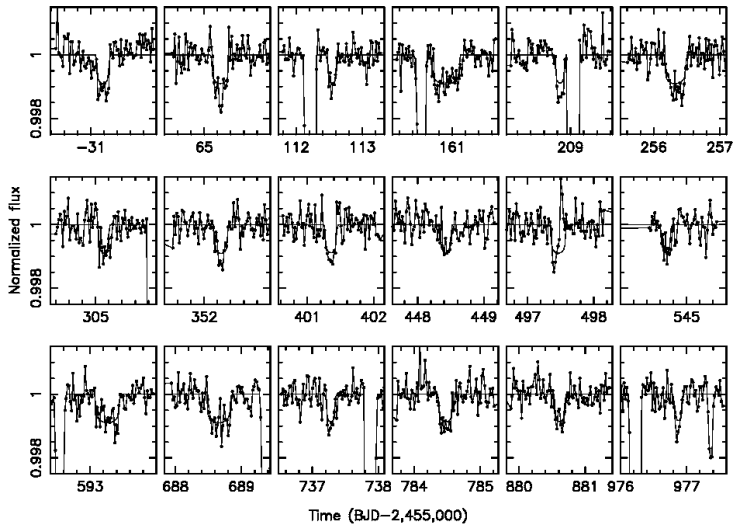} 
 \caption{
Each of the eighteen Kepler-47 b transits and the  
photometric-dynamical model fits. The occasional large 
downward excursions are due to eclipses.
}
   \label{fig2}
\end{center}
\end{figure}

\begin{figure}[!ht]
\begin{center}
 \includegraphics[width=4.0in]{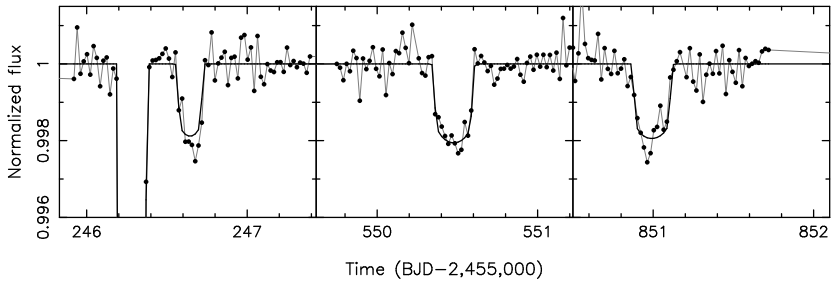} 
\vspace*{1.0 cm}
 \caption{
The three observed Kepler-47 c transits with model fits. The
large excursion in the first panel is a primary eclipse.
}
   \label{fig3}
\end{center}
\end{figure}

The orbit of the binary has a very mild eccentricity ($e=0.023$) 
and a period of 7.45 days. The orbital periods of the planets
are 49.5 d and 303.1 d, respectively (Kepler-47c is currently
the longest period confirmed transiting planet). The eccentricity 
of planet
b is low ($e<0.035$). For planet c, only 3 transits are measured so
the eccentricity is harder to constrain; a 95\% upper limit of 
$e<0.41$ is found, with $e\sim 0.1$ preferred. The period of planet b 
is 6.6 times the binary period, 77\% larger than the critical period
below which interaction with the binary stars could lead to a 
dynamical instability 
(\cite[Holman \& Wiegert 1999]{Holman1999}).
While planet b is well interior to the habitable zone (HZ), 
planet c lies completely within the HZ for $e<0.2$, and even with
$e=0.4$ the mean insolation is 96\% of the Sun-Earth insolation.
It is the mean insolation that is relevant for habitability
(\cite[Williams \& Pollard 2002]{Williams2002}).

In addition to the transits caused by planets b and c, there is an
unexplained single transit-like event of depth 0.2\%. The signal is 
six cadences wide and is not an obvious instrumental artifact.
It is tempting to suspect an additional planet as the source of this 
``orphan'' transit, but with only one event measured, the parameter
space is too unconstrained to place confidence in the planet 
hypothesis -- though we note that dynamically there is plenty of 
room between planets b and c for a third planet.

Kepler-47 is significant because
it establishes that planetary {\it systems} (not just single planets)
can form {\it and persist} in the chaotic environment close to a 
binary star. 
The binary stars tend to augment planet-planet interactions 
(\cite[Youdin et al.\ 2012]{Youdin2012}), so planetary orbits that 
lie close to the binary are highly susceptible to dynamical 
instability. And Kepler-47c shows that circumbinary planets can 
exist in the habitable zones around their host stars, opening up 
new habitats for life to potentially exist.

\begin{figure}[ht!]
\begin{center}
 \includegraphics[width=3.4in]{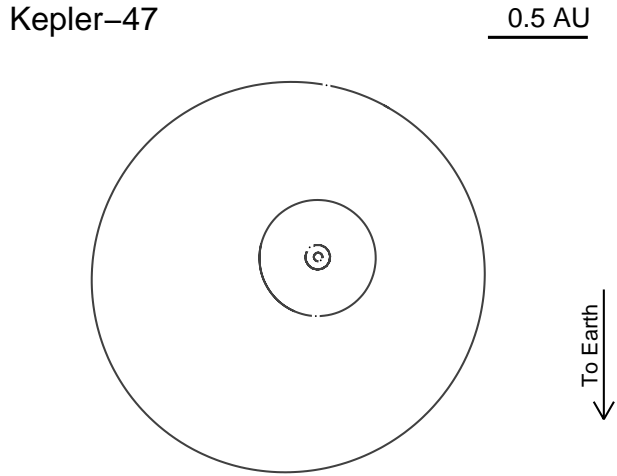} 
 \caption{
Bird's eye view of the binary and planetary orbits for Kepler-47.
The small gaps in the orbit curves show the location of the bodies 
at this particular epoch.
}
   \label{fig4}
\end{center}
\end{figure}

\section{The Developing Circumbinary Planet Picture}
The stars in Kepler-16 have an eccentricity of 0.159 and star A is 
3.4 times the mass of star B. This sizable mass ratio means star A 
orbits relatively close to the barycenter of the binary. When only 
Kepler-16 was known, one might have speculated that these stars 
are in a particularly benign configuration for planet formation and
migration. But Kepler-34 and 35 quickly refuted that notion 
because the stars are close to equal mass (ratio A/B is 1.03 and 1.10  
respectively), and so both stars orbit the barycenter nearly equally.
Furthermore, Kepler-34 A and B have highly eccentric orbits
($e$=0.521) 
and their interaction with the planet is strong enough
to produce a noticeable deviation from a Keplerian solution even 
after just one period.
It is now clear that there is a wide range of stellar configurations
for which circumbinary planets can exist.
Primary star masses range from 0.69 to 1.53 M$_{\odot}$ 
(Kepler-16 A \& PH1 Aa), 
mass ratios from 1.03 to 3.76 (Kepler-34 \& PH1),
and eccentricity from 0.023 to 0.521 (Kepler-47 \& Kepler-34).
Likewise, the planetary orbits have a sizeable spread in 
eccentricities, ranging from nearly circular e=0.007 to
a significant $e$=0.182 (Kepler-16 \& Kepler-34).
There is no tendency for orbital resonances with the binary. 
It is clear that no special geometry is favored, with two notable 
exceptions: co-planarity and close-in orbits.

All seven {\it Kepler} circumbinary planets orbit their stars very 
close to the plane of the binary (in a prograde direction). And in 
cases where the stellar spin axis has been measured (Kepler 16 and 47; 
and possibly PH1 based on the match between the observed and expected 
$V_{rot}\sin{i}$), the spin axes are closely aligned with the binary 
axis. While tidal forces act to align the spin axes with the orbital 
axis (e.g. see \cite[Winn, et al.\ 2011]{Winn2011}),
the planet and stellar orbital co-planarity suggests that a single-disk 
formation and migration scenario is likely.

The propensity of circumbinary planets to orbit close to their host 
stars has been noted since the discovery of Kepler-16.
The (inner) planets orbit surprisingly close to the boundary where 
dynamical instabilities due to perturbations from the binary can be 
present (\cite[Holman \& Wiegert 1999]{Holman1999}).
The observed circumbinary planets have semi-major axes that lie 
between 1.09 and 1.46 times the 
critical radius. The reason for this is unclear.
Migration might become inefficient near the critical radius,
leaving planets just outside this radius. Or migration may operate
normally but any planets within the critical radius are lost, leaving
only those at larger radii. Or this is just an observational bias - the 
closer in the planet, the more likely it will transit the stars.
Another interesting observation is that the circumbinary planets 
tend {\it not} to exist around the shorter period {\it Kepler} 
eclipsing binaries.
The shortest period {\it Kepler} eclipsing binary hosting a planet
is 7.44~d (Kepler-47; the longest is Kepler-16 with a binary period of 
41~d). While a strong observational bias against detecting long period 
planets is present, the opposite is true for short period binaries
(assuming the tendency for the planet to orbit near the critical 
radius continues to hold). And in addition to the geometrical factors,
the shorter period binaries allow shorter period planets, so more
transits should be present in the {\it Kepler} data allowing easier
detection. The majority of {\it Kepler} eclipsing binaries have periods 
less than 1 day. Since these short-period binaries are unlikely to have 
formed in such a tight orbit, their lack of planets may be related to 
the mechanism that removed angular momentum and allowed the stars to 
orbit so closely. The apparent lack of circumbinary planets around 
short-period binaries is worthy of investigation. 

An interesting characteristic of the circumbinary planets is that all 
eight have a size (mass and/or radius) smaller than Jupiter. Since a 
larger-size planet is more likely to be found than a smaller planet, 
this cannot be a selection effect.
\cite[Pierens \& Nelson (2008)]{Pierens2008} predicted such a tendency,
based on simulations of the orbital evolution of planets embedded in
a circumbinary disk. This certainly deserves attention, especially if 
the trend continues as more circumbinary planets are discovered.

Finally, the tendency for the observed circumbinary planets to lie 
close to their
critical radius has an interesting consequence for habitability.
The stability criterion requires the planet to orbit outside roughly 
$\sim$2-4 times the binary semi-major axis, or periods $\sim$3-8 times 
the binary period. The known circumbinary planets
have binaries with periods roughly around 
$\sim$10-40 days, so the planets will have periods a bit larger than 
$\sim$30-320 days. Since the {\it Kepler} targets are mostly G and K 
type stars, this orbital period is rather close to the habitable
zone (HZ) around the binary. The first {\it Kepler} circumbinary 
planet discovered was just exterior to its HZ and thus 
too cold, while the second circumbinary planet discovered was 
slightly too hot. While extrapolating from two cases is
inherently foolish, such sophistry did bear fruit with Kepler-47c --
it is quite remarkable that it only took the discovery of five 
circumbinary systems to find a planet squarely in the HZ.
Table 1 lists all the known transiting circumbinary planets as of 2012, 
with some of their characteristics related to the HZ. Of course none 
of the planets are terrestrial, but large moons of planets 
in the HZ would be very interesting. As the {\it Kepler} data are
searched with better methods, smaller circumbinary planets will be 
teased from the data, and it will not be surprising to find an 
Earth-like circumbinary planet in the HZ. Because of the stellar 
binarity, the insolation received by the planet will likely be 
time-varying in an interesting way, quite unlike the steady insolation 
Earth receives from the Sun.
The observational study of ``tatooines'' has just begun, and already
we have found a system with at least two planets, a planet in the HZ, 
systems with rich dynamics, clues on planet and star formation/migration, 
and of course, abundant puzzles to nourish the human mind.

\begin{table}[h]
  \begin{center}
  \caption{Kepler Transiting Circumbinary Planets}
  \label{tab1}
 {\scriptsize
  \begin{tabular}{|c|c|c|c|c|c|}\hline 
Planet & P$_{\rm{planet}}$ &  P/P$_{\rm{crit}}$ for &  
  Insolation $<$S$>$  &  T$_{\rm{equilibrum}}$$^{1}$  & 
Habitable Zone\\ 
       &      (days)       &  (in)stability         &
  (Sun-Earth)         &                           &  proximity  \\ 
\hline
Kepler 16 b &  229 &  1.14 & 0.3 & 180 K / -93 C & little too 
cold$^{2}$\\ \hline
Kepler-34 b &  289 & 1.21  & 2.4  & 312 K / 39 C  & little too hot \\ 
\hline
Kepler-35 b &  131 & 1.24  & 3.6  & 345 K / 72 C  & too hot     \\ 
\hline
Kepler-38 b &  105 & 1.42  & 12.8 & 475 K / 202 C & way too hot \\ 
\hline
Kepler-47 b &  49.5 & 1.77 & 9.6  & 442 K / 169 C & way too hot \\ 
\hline
Kepler-47 c &  303 &  10.8 & 0.88 & 243 K / -30 C  & Goldilocks! \\ 
\hline
PH1-b       &  138 &  1.29 & 10.6  & 454 K / 180 C  & way too hot \\ 
\hline
  \end{tabular}
  }
 \end{center}
\vspace{1mm}
\hspace{12mm}
 \scriptsize{
  $^1$Assuming a Bond albedo of A$_{\rm B}$ = 0.34.
\hspace*{0.3cm}
  $^2$But it does lie in the {\it extended} HZ.}
\end{table}

\vspace*{8mm}
Acknowledgments:\\
Kepler was selected as the 10th mission of the Discovery Program,
with funding provided by NASA, Science Mission Directorate. W.F.W. 
and J.A.O. gratefully acknowledge support from the Kepler 
Participating Scientist Program via NASA grant NNX12AD23G
and from the NSF via grant AST-1109928.
J.A.C. and D.C.F. acknowledge NASA support through Hubble Fellowship 
grants, awarded by STScI, operated by AURA.


\end{document}